\documentclass[aps,prl,twocolumn,showpacs]{revtex4}
\usepackage{graphicx}
\usepackage{amssymb}
\usepackage{acronym}
\usepackage{amsmath}
\usepackage{amsfonts}%
\begin{document}
\title{ Spin-polarized surface states close to adatoms on Cu(111)}
\author{B. Lazarovits$^{1}$, L. Szunyogh$^{1,2}$, P. Weinberger$^{1}$}
\affiliation{$^1$Center for Computational Materials Science,
Technical University Vienna,\\
A-1060, Gumpendorferstr. 1.a., Vienna, Austria \\
$^2$Department of Theoretical Physics and  \\
Center for Applied Mathematics and Computational Physics,\\
Budapest University of Technology and Economics, \\
Budafoki \'{u}t 8., H-1521 Budapest, Hungary}

\date{\today}

\begin{abstract}
We present a theoretical study of surface states close to \textit{3d}
transition metal adatoms (Cr, Mn, Fe, Co, Ni and Cu) on a Cu(111) surface in
terms of an embedding technique using the fully relativistic
Korringa-Kohn-Rostoker method. For each of the adatoms we found resonances in
the \textit{s}--like states to be attributed to a localization of the surface
states in the presence of an impurity. We studied the change of the
\textit{s}--like densities of states in the vicinity of the surface state
band-edge due to scattering effects mediated via the adatom's \textit{d}%
-orbitals. The obtained results show that a magnetic impurity causes
spin-polarization of the surface states. In particular, the long-range
oscillations of the spin-polarized \textit{s}--like density of states around an
Fe adatom are demonstrated.

\end{abstract}
\maketitle

\section{Introduction}

Following the first experimental observation of a surface band at
Cu(111) in terms of angle-resolved photoemission (ARPES) \cite{GS75} 
electrons at closed packed surfaces of noble metals
have been at the center of much experimental \cite{K83,CLE+95,CLE+96,MLE00,BGO01,SPT+04}
and theoretical \cite{HP94,Li+98,SNL+04} attention.
For a pristine surface the energies of Shockley surface
states lie in the 'gap' around the L-point of the bulk Brillouin Zone, their
wavefunctions being confined to the surface. The corresponding dispersion
relations ave been determined by ARPES
and were found to be two-dimensional free--electron like parabolas
\cite{GS75,K83,BGO01}. 
One of the highly interesting features of this phenomenon is
the response to perturbations causes by placing, e.g., adatoms on the surface.
As to be expected such a response is characterized by long range Friedel like
charge oscillations governed by the dispersion relation of the two-dimensional
surface electron gas. The existence of long range interactions between adatoms
on surfaces supporting surface states can lead to the formation of an atomic
superlattice as recently shown both experimentally\cite{SPT+04} and
theoretically \cite{SNL+04}. Until recently STM studies of atoms on well
defined Cu, Ag and Au (111) surfaces imaged only the charge distribution of
the surface electrons \cite{CLE+95,CLE+96,MLE00}. Remarkable developments in
spin polarized STM \cite{B03}, however, are expected to detect spatial
variations in the magnetic density, which in turn might provide an
understanding of the magnetic behavior of this kind of systems
\cite{HCB02,CFR+05,LUS+05,SNH+05}. Evidently, this also opens up the possibility of
designing magnetic nano-structures for both scientific and technological purposes.

In accordance with a theoretical prediction by Borisov \textit{et al.}
\cite{BKG02}, by using STM at 7 K, Limot \textit{et al.} \cite{LPK+05} very
recently reported on an adatom-induced localization of the surface states
electrons on Cu(111) and Ag(111) surfaces. The existence of such resonances
for a single Cu adatom was also shown theoretically by Olsson \textit{et al.}
\cite{OPB+04} in terms of a parameter-free pseudopotential method. The
appearance of a peak in the density of states (DOS) just below the surface
state band can be attributed to a theorem by Simon claiming the existence of a
bound state for any attractive potential in two dimensions~\cite{S76,OPB+04}.
It was demonstrated in Ref.~\cite{LPK+05} by comparing results for Co and Cu
adatoms on Cu(111) that the type of adsorbate influences the shape of the
adatom-induced resonance. For this very reason we performed a systematic study
for a series of \textit{3d} impurities (Cr, Mn, Fe, Co, Ni and Cu) on Cu(111)
in order to identify spin-dependent features of the occurring resonance.

\section{Method of calculations}

\label{sec:theo}

Within multiple scattering theory the electronic structure of an ensemble of
non-overlapping 
potentials is described by the so-called scattering
path operator (SPO) matrix (for more details see, e.g., Ref. \cite{weinbook2}).
The SPO matrix $\mbox{\boldmath $\tau$}_{\mathcal{C}}$ that refers to a
finite cluster $\mathcal{C}$ embedded into a host system can be obtained from
the following Dyson equation \cite{LSW02},%
\begin{equation}
\mbox{\boldmath $\tau$}_{\mathcal{C}}(E)=\mbox{\boldmath $\tau$}_{h}(E)\left[
\mbox{\boldmath $I$}-(\mbox{\boldmath $t$}_{h}^{-1}%
(E)-\mbox{\boldmath $t$}_{\mathcal{C}}^{-1}(E))\mbox{\boldmath $\tau$}_{h}%
(E)\right]  ^{-1} \; , \label{eq:dyson}%
\end{equation}
where $\mbox{\boldmath $t$}_{h}(E)$ and $\mbox{\boldmath $\tau$}_{h}(E)$
denote the single--site scattering matrix and the SPO matrix for the
unperturbed host confined to sites in $\mathcal{C}$, respectively, while
$\mbox{\boldmath
$t$}_{\mathcal{C}}$ denotes the single--site scattering matrices of the
embedded atoms. Note, that Eq.~(\ref{eq:dyson}) takes into account all
scattering events both inside and outside the cluster.

First, a fully self--consistent (SC) calculation is performed for the Cu(111)
surface by means of the screened Korringa--Kohn--Rostoker (SKKR) method
\cite{SUW95}. Then for single adatoms placed on top of Cu(111) the multiple
scattering problem is solved self--consistently in terms of the embedding
method discussed in details in Ref. \cite{LSW02}. 
The selfconsistent calculations for both the semi-infinite Cu(111)
host and the adatoms were performed using the atomic sphere approximation
(ASA) and the local spin--density approximation in the parameterization of
Vosko \emph{et al.} \cite{VWN80}. Due to the fully relativistic treatment
applied the orientation of the effective magnetic field had to be specified:
it was chosen to point along the $z$ axis (normal to the surface). In order to
evaluate the inevitable Brillouin-zone integrations, 
in the selfconsistent calculations 70
$k_{\parallel}$--points were used in the irreducible wedge of the surface
Brillouin-zone (SBZ). For the calculation of the $t$--matrices and for the
multipole expansion of the charge densities, necessary to evaluate the
Madelung potentials, a cut--off of $\ell_{max}=2$ was assumed. Energy
integrations were performed by sampling 16 points on a semicircular contour in
the complex energy plane according to an asymmetric Gaussian quadrature. 
The densities of states were
calculated parallel to the real energy axis
with an imaginary part of 5~mRyd by sampling
about 2200 $k_{\parallel}$ points within the irreducible wedge of the SBZ.
In the present calculations no attempt was made to include
surface relaxations: the geometry was taken to be identical to an ideal Cu
bulk fcc lattice (lattice constant $a_0=3.614$~\AA).

\section{Results and discussion}

\begin{figure}[th] 
\includegraphics[width=0.45\textwidth,clip]{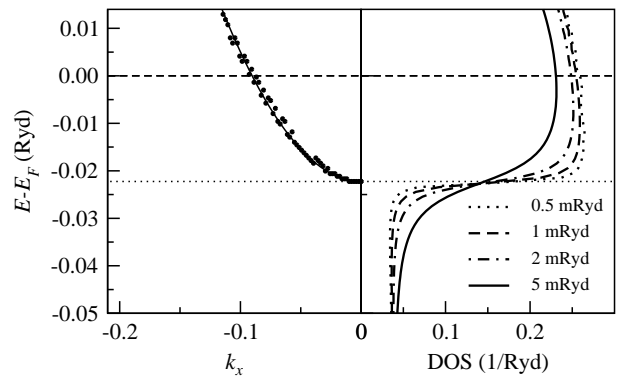}
\caption{ \textbf{Left panel:} Positions of maxima in the Bloch spectral
functions  near to the $\bar{\Gamma}$ point of the SBZ (dots). The solid line
refers to a parabolic  fit. It should be noted that only the first third of
the SBZ is displayed ($\bar{K}\approx0.65$). \textbf{Right panel:} Density of
the surface states ($|k_{\parallel}|<0.15$) close to a clean Cu(111) surface. 
The different values for the imaginary part of the energy used 
are displayed explicitely.
The Fermi energy and the bottom of
the surface state band are indicated by a dashed and a dotted 
horizontal lines, respectively.}
\label{fig:band}
\end{figure}

In order to determine the dispersion relation and the effective mass of the
surface electrons \textit{Bloch-spectral functions} (BSF) \cite{FS80} were
evaluated in terms of the SKKR method for k-points between the $\bar{\Gamma}$
and $\bar{K}$ points in the fcc(111) SBZ close to the Fermi energy. It should
be recalled that the dispersion relation of the surface state band can be
defined by the position of the maxima in the BSF. In agreement with
experiments the calculated dispersion relation is a free--electron like
parabola as indicated in Fig.~\ref{fig:band}. The bottom of the calculated
surface state band is only about 0.3~eV below the Fermi energy which is a bit
smaller than the experimental values (0.4~eV) \cite{K83,BGO01}. Also in good
agreement with the experimental data \cite{K83} is the corresponding effective
mass $m^{\ast}=0.37\,m_{e}$ as obtained using an appropriate fitting procedure.
In Fig.~\ref{fig:band} also the density of the surface states in the vicinity 
of the band-edge is displayed.
Shown is the {\em s}-like DOS of the first vacuum (empty sphere) layer
as integrated over a sphere of a radius of 0.15/a.u. centered around
$\bar{\Gamma}$ of the SBZ.  Note that the corresponding DOS at the substrate
layers decays exponentially with increasing distance from the surface.
As our calculations do not include structural defects (e.g., steps)
at the surface, electron-electron interaction beyond the density functional
theory or electron-phonon interaction, 
the rather broad onset of the surface-state band
is a direct consequence of using an imaginary part of 5 mRyd 
for the energy when calculating the DOS,
see the solid line on the right of Fig.~\ref{fig:band}.
In order to justify this argument, only for this particular case,
we also performed calculations with imaginary parts of 2, 1 and 0.5 mRyd
displayed in Fig.~\ref{fig:band} in terms of dash-dotted, dashed and dotted
lines, respectively.
As can be seen, by decreasing the imaginary part of the energy the onset
of the surface-band approaches a step-like behavior. The experimental onset
($\sim$~30 meV \cite{LPK+05})
is fairly well recovered in case of 1 mRyd for the imaginary part.
It should be noted that, in order to smooth out spurious oscillations 
in the calculated 
density of the surface states, in this case a sampling of more 
then 50~000 $k_{\parallel}$ points  
within the irreducible wedge of the SBZ 
(about 2000 $k_{\parallel}$ points for $|k_{\parallel}|<0.15$) was 
needed, 
see also a recent theoretical STM study of Hofer and Garcia-Lekue 
\cite{HG05}. 
Fortunately, however, 
applying a Lorentzian broadening of 5 mRyd ($\sim$ 70 meV) 
for the DOS turned out to be sufficient to resolve the adatom induced 
surface states to be studied in this paper.



\begin{figure}[th]
\includegraphics[width=0.48\textwidth,clip]{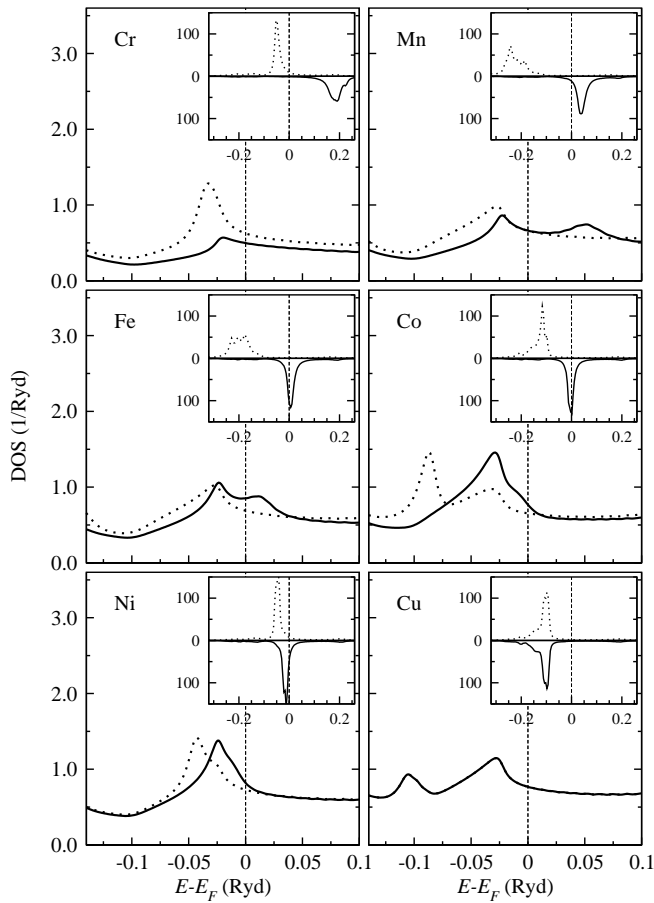}\caption{ Calculated
spin-projected \textit{s}--like density of states at the impurity positions.
Insets: Spin-projected total density of states of the adatoms. In each entry,
the majority and the minority spin DOS's are depicted by dotted and solid
lines, respectively. }%
\label{fig:dos_all}%
\end{figure}

In Fig.~\ref{fig:dos_all} the calculated \textit{s}--DOS and total DOS
(insets) for the chosen series of adatoms are presented as projections with
respect to the two spin directions. As is also evident from the spin--split
peaks in the total DOS, dominated mainly by \textit{d}--like contributions,
the Cr, Mn, Fe, Co and Ni adatoms on Cu(111) were found to be magnetic with
spin--moments of $S_{\mathrm{Cr}}^{z}=4.21\,\mu_{B}$, $S_{\mathrm{Mn}}%
^{z}=4.39\,\mu_{B}$, $S_{\mathrm{Fe}}^{z}=3.27\,\mu_{B}$, $S_{\mathrm{Co}}%
^{z}=2.02\,\mu_{B}$ and $S_{\mathrm{Ni}}^{z}=0.51\,\mu_{B}$, while the Cu
adatom is non-magnetic. It should be noted that within a relativistic
description the electronic spin is not a constant of motion. In the present
cases, however, characterized by large exchange splittings and weak spin-orbit
interactions it is quite illustrative to view the two spin--channels separately
by making use of the approximation discussed in Ref. \cite{SGP+88}.


Concentrating first on the case of the Cu adatom, quite a broad peak is found
just below the bottom of the surface state band ($E\approx-0.025$~Ryd). The
appearance of this peak indicates that adatoms act as an effective attractive
potential for surface state electrons inducing thus a certain degree of
localization. The widths and positions with respect to the surface state band
edge of the adatom-induced peak agrees well with the one obtained in
Ref.~\cite{OPB+04}. 
The calculated resonance width being also in agreement with
Ref.~\cite{OPB+04} is, however, approximately three times larger than 
experimentally measured \cite{LPK+05}. 
This disagreement with the
experiment can most probably be attributed again to the rather 
large Lorentzian broadening of the calculated DOS. 

\begin{figure}[th]
\includegraphics[width=0.4\textwidth,clip]{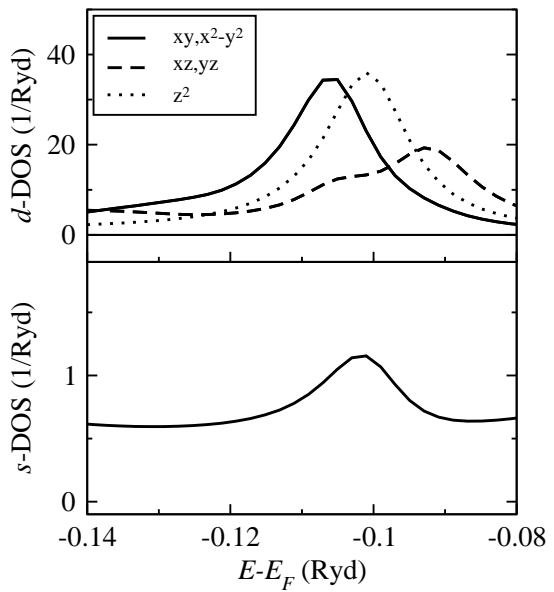}\caption{ \textbf{Upper
panel}: Calculated $d$--DOS of a single Cu impurity on a Cu(111) surface
as partitioned into  $d_{xz}$, $d_{yz}$, $d_{xy}$, $d_{z^{2}}$ and
$d_{x^{2}-y^{2}}$ like contributions. \textbf{Lower panel}: $s$--DOS at
the same site. The DOS's presented in this figure correspond to one of the
spin--projections. }%
\label{fig:cudos}%
\end{figure}

The fairly narrow \textit{d}-band of the Cu adatom lies well below the 2D
surface band ($E_{d}\approx-0.1$~Ryd). It can therefore be assumed that the
adatom-induced resonance at $E\approx-0.025$~Ryd is hardly influenced by
\textit{d}--like states. This allows us to consider the Cu adatom as a
reference when identifying effects of \textit{d}-states on the adatom-induced
resonance in the case of other impurities. In line with Ref.~\cite{OPB+04}, a
peak in the \textit{s}--DOS of the Cu adatom can be observed
just at the peak position of the \textit{d}-band. 
This is a direct consequence of the
hybridization between \textit{s} and \textit{d} type atomic orbitals as
illustrated in Fig.~\ref{fig:cudos} using group theoretical arguments. It is
important to note that in order to make use of a well-defined
non-relativistic classification of the densities of states in terms of
symmetry adapted spherical harmonics, only in this case we "switched off" the
spin-orbit coupling by applying the so-called scaling scheme proposed by Ebert
\textit{et al.} \cite{EFV+96}. In the upper panel of this figure the
\textit{d}--DOS is decomposed into three components: two of them corresponding
to the two--dimensional irreducible representation ($E$) of the $C_{3v}$
point--group, namely, according to the following sets of symmetrized basis 
functions $\{d_{xy},d_{x^{2}-y^{2}}\}$ and $\{d_{xz},d_{yz}\}$,
and one to a one-dimensional representation ($A_{1}$), namely, with 
respect to $\{d_{z^2} \}$. Clearly enough, the \textit{s}-states that correspond to $A_{1}$ symmetry
can hybridize only with $d_{z^{2}}$--states as is apparent from the
line--shapes of the corresponding DOS's in Fig.~\ref{fig:cudos}.


From Fig.~\ref{fig:dos_all} the appearance of the adatom-induced localized
state is also obvious in the case of the magnetic impurities as only a slight
variation of its position in energy occurs. This indicates that each kind of
impurity acts in a similar manner as an attractive potential well for the 2D
electron gas. As can be seen, however, in the case of magnetic adatoms the
shape of the adatom-induced resonance  is different for the two spin
projections. This observation can be well explained in terms of  the energetic
position of the corresponding $d$-bands of the impurities. For Mn, Fe and also for Co the
majority \textit{d}-band lies reasonably deep below the adatom-induced
resonance. Therefore, the line shape of 
one (say majority) spin--projected 
\textit{s}--DOS around $E=-0.025$ Ryd is practically unchanged as
compared to the Cu case. Within the energy range displayed, an 
$s$-$d$ hybridized peak in the majority $s$--DOS is seen for Co
around -0.09 Ryd. An analysis in terms of  symmetry-adapted spherical
harmonics (basis functions)  as discussed in the case of the Cu adatom applies
also in this case. For Cr and Ni the position of the majority \textit{d}-band
is just at the bottom of surface band. The adatom-induced resonance appears
therefore for Ni  just as a small shoulder on the upper side of the $s$-$d$
hybridized peak, while for Cr the resonance can hardly be traced at all.


For all magnetic adatoms investigated the minority $d$-band overlaps with the
2D surface state band. In the series Cr to Ni its position moves downwards
in energy, its width decreases monotonously. For Mn and Fe a
moderately well-developed  peak  can be seen at the position of the very sharp
$d$--band, while for Co and Ni just a small 'hump' in the minority $s$--DOS
\ is visible. The occurrence of these structures is of similar origin as
the $s$-$d$ hybridized peaks in the majority spin channel: the continuum-states
experience resonant scattering due to an overlap with an adatom's $d$ 
orbitals giving thus rise to well--known virtual bound state resonances. 
As can be shown in terms of a non-interacting Anderson model,
this effect is proportional to the DOS of the continuum band, which
decreases if the position of the $d$-band moves towards the bottom of the
surface band. This explains qualitatively the trend  for the width of the
minority $d$-band. It is, however, apparent that the minority spin
adatom-induced resonance peak ($E\approx-0.025$ Ryd) is considerably larger
for Co and Ni as for Cu, a fact which might be attributed to a
strengthening of the localization of the 2D surface electrons due to
$d$--like virtual bound states very close in energy.

\begin{figure}[th]
\includegraphics[width=0.35\textwidth,clip]{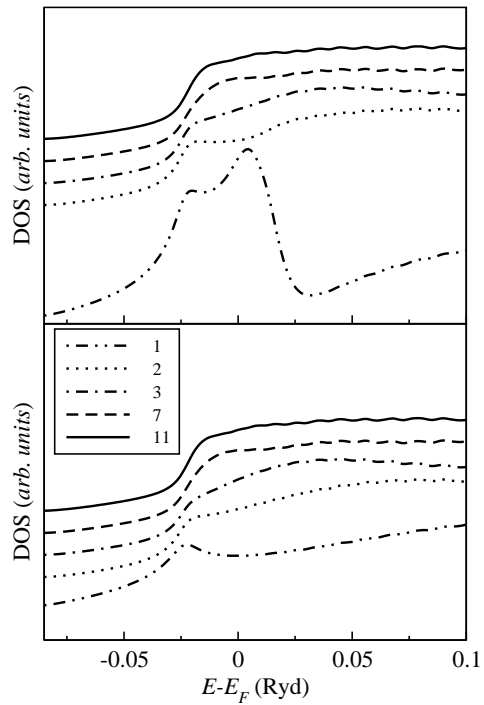}\caption{ Calculated
minority spin (upper panel) and majority spin (lower panel) \textit{s}--DOS's 
at a site in the same plane as a single Fe impurity on a Cu(111) surface
with respect to the distance between these two sites. The distance (in units
of the two--dimensional lattice constant)  is indicated in the legend of the
lower panel. }%
\label{fig:spatdos}%
\end{figure}

In Fig.~\ref{fig:spatdos} the \textit{s}--DOS for both spin directions is
shown for a site at selected distances measured with respect to an Fe
adatom in the same plane (parallel to the surface). It can be observed that
the line shapes related to the adatom-induced resonance and the minority spin
resonant scattering vanish in fact at a radius of about two in units of the 
2D (in plane) lattice constant. 
Beyond this radius a lineshape characteristic for the 2D
surface--state band appears to evolve, however, with a broader onset than for
the pristine Cu(111) surface. This broadening can again be attributed to the
interaction (overlap) between the surface states and the
$d$-states of an adatom.


\begin{figure}[th]
\includegraphics[width=0.4\textwidth,clip]{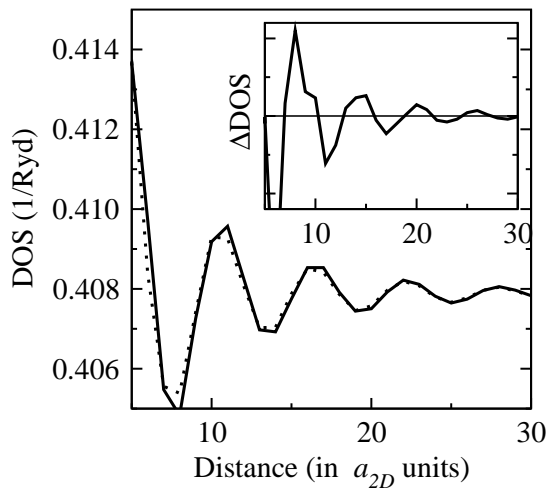}\caption{Calculated
minority spin (solid line) and majority spin (dotted line) $s$--DOS's at
$E-E_{F}=0.012$~Ryd  of a site positioned in the same plane but at a certain
distance  from a single Fe impurity on a Cu(111) surface. Inset: difference of
the spin projected $s$--DOS's. }%
\label{fig:poldos}%
\end{figure}

At a distance of seven 2D lattice constants ($\sim17.9$~\AA ) the $s$--DOS is practically
the same as calculated for the clean Cu(111) surface. A long--ranged
oscillatory behavior of the densities of states can, however, still be
resolved. In Fig.~\ref{fig:poldos} the DOS at a selected energy, namely,
34~mRyd above the bottom of the surface--state band is displayed  for both
spin channels as a function of the distance  from an Fe adatom. A simple
estimate of the wavelength of the 2D Friedel oscillation,
\begin{equation}
\lambda=\frac{\pi}{\hbar\sqrt{2m^{\ast}E}}\;,
\end{equation}
gives $\lambda\approx15$~\AA ~$\approx6\,a_{2D}$ that can be read off \ from
Fig.~\ref{fig:poldos}. As shown in the inset of Fig.~\ref{fig:poldos}, a
magnetic impurity also induces long--range oscillations in the magnetization
density of states (MDOS, defined as the difference of the spin--projected
DOS's), with the same period. Clearly, these oscillations in the MDOS lead
to a long--range RKKY interaction on noble metal (111) surfaces as  discussed,
e.g.,  in Ref.~\cite{SNL+04}.


\section{Conclusion}

We presented a series of calculations for spin-dependent surface states close
to \textit{3d} transition metal adatoms (Cr, Mn, Fe, Co, Ni and Cu) on a
Cu(111) surface. In agreement with previous theoretical \cite{OPB+04} and
experimental \cite{LPK+05} studies the appearance of an adatom-induced resonance
just below the surface state band edge is shown. The occurrence of an
additional peak in the $s$--DOS caused by the adatom's $d_{z^{2}}$-states
was found below the adatom-induced resonance not only for a Cu
adatom~\cite{OPB+04} but also for  magnetic adatoms in the majority spin
channel. We also found evidence of  an interaction of the minority spin
$d$-states of the magnetic impurities and the surface state continuum, that
for Co and Ni possibly  explains the remarkable enhancement of the
adatom-induced resonance peak in the minority spin channel. The different
shape of the adatom-induced resonance for Cu and Co was indeed observed in
experiments \cite{LPK+05} and is in qualitative agreement with our
calculations. Finally, we pointed out the existence of long--range
spin--polarized oscillations of the surface states around a magnetic impurity
being the very origin of a 2D RKKY interaction.

\acknowledgments Financial support was provided by the Center for
Computational Materials Science (Contract No. GZ 45.531, GZ 98.366), the
Austrian Science Foundation (Contract No. W004), the Hungarian Research and
Technological Innovation Council and the Bundesministerium f\"{u}r Auswertige
Angelegenheiten of Austria (Contract No. A-3/03) and the Hungarian National
Scientific Research Foundation (OTKA T037856 and T046267).


\begin{thebibliography}{99}

\bibitem{GS75}
P. O. Gartland and B. J. Slagsvold, Phys. Rev. B {\bf 12}, 4047 (1975).

\bibitem{K83}
S.~D. Kevan, Phys. Rev. Lett. {\bf 50},  526  (1983).

\bibitem{CLE+95}
M.~F. Crommie, C.~P. Lutz, D.~M. Eigler, and E.~J. Heller, Physica D {\bf 83},
  98  (1995).

\bibitem{CLE+96}
M.~F. Crommie, C.~P. Lutz, D.~M. Eigler, and E.~J. Heller, Surface Science {\bf
  361/362},  864  (1996).

\bibitem{MLE00}
H.~C. Manoharan, C.~P. Lutz, and D.~M. Eigler, Nature {\bf 403},  512  (2000).

\bibitem{BGO01}
F. Baumberger, T. Greber, and J. Osterwalder, Phys. Rev. B {\bf 64},  195411
  (2001).

\bibitem{SPT+04}
F. Silly, M. Pivetta, M. Ternes, F. Patthey, J. P. Pelz, and W.-D. Schneider, 
Phys. Rev. Lett. {\bf 92},  016101  (2004).

\bibitem{HP94}
G. H\"ormandinger and J.~B. Pendry, Phys. Rev. B {\bf 50},  18607  (1994).

\bibitem{Li+98}
J. Li, W.-D. Schneider, R. Berndt, O. R. Bryant, and S. Crampin,
Phys. Rev. Lett. {\bf 81}, 4464 (1998).

\bibitem{SNL+04}
V. S. Stepanyuk, L. Niebergall, R. C. Longo, W. Hergert, and P. Bruno, 
Phys. Rev. B {\bf 70},  075414  (2004).

\bibitem{B03}
M. Bode, Rep. Prog. Phys. {\bf 65},  523  (2003).

\bibitem{HCB02}
K. Hallberg, A.~A. Correa, and C.~A. Balseiro, Phys. Rev. Lett. {\bf 88},
  066802  (2002).

\bibitem{CFR+05}
A.~A. Correa, F.~A. Reboredo, and C.~A. Balseiro, Phys. Rev. B {\bf 71},
  035418  (2005).

\bibitem{LUS+05}
B. Lazarovits, B. \'Ujfalussy, L. Szunyogh, B. L. Gy\"orffy, P. Weinberger, 
J. Phys.: Condens. Matter {\bf 17},  S1037  (2005).

\bibitem{SNH+05}
V. S. Stepanyuk, L. Niebergall, W. Hergert, and P. Bruno, Phys. Rev. Lett. {\bf 94}, 187201 (2005).

\bibitem{BKG02}
A.~G. Borisov, A.~K. Kazansky, and J.~P. Gauyacq, Phys. Rev. B {\bf 65},
  205414  (2002).

\bibitem{LPK+05}
L. Limot, E. Pehlke, J. Kr\"oger, and R. Berndt, Phys. Rev. Lett. {\bf 94},
  036805  (2005).

\bibitem{OPB+04}
F.~E. Olsson, M. Persson, A.G. Borisov, J.-P. Gauyacq, J. Lagoutei, and 
S. F\"olsch, Phys. Rev. Lett. {\bf 93},  206803  (2004).

\bibitem{S76}
B. Simon, Ann. Phys (N.Y.) {\bf 97},  279  (1976).

\bibitem{weinbook2}
J. Zabloudil, R. Hammerling, L. Szunyogh, and P. Weinberger, {\em Electron
  Scattering in Solid Matter} (Springer-Verlag, Berlin Heidelberg, 2005).

\bibitem{LSW02}
B. Lazarovits, L. Szunyogh, and P. Weinberger, Phys. Rev. B {\bf 65},  104441
  (2002).

\bibitem{SUW95}
L. Szunyogh, B. \'Ujfalussy, and P. Weinberger, Phys. Rev. B {\bf 51},  9552
  (1995).

\bibitem{VWN80}
S.~H. Vosko, L. Wilk, and M. Nusair, Can. J. Phys. {\bf 58},  1200  (1980).

\bibitem{FS80}
J.~S. Faulkner and G.~M. Stocks, Phys. Rev. B {\bf 21},  3222  (1980).

\bibitem{HG05}
W. A. Hofer and A. Garcia-Lekue, Phys. Rev. B {\bf 71}, 085401 (2005)

\bibitem{SGP+88}
J.~B. Staunton, B.~L. Gy\"orffy, J. Poulter, and P. Strange, J. Phys. C {\bf
  21},  1595  (1988).

\bibitem{EFV+96}
H. Ebert, H. Freyer, A. Vernes, and G. Guo, Phys. Rev. B {\bf 53},  7721
  (1996).

\end{thebibliography}

\end{document}